\documentclass{svproc}
\usepackage{url}

\usepackage{amsmath,amssymb,bbm,enumerate}

\usepackage[all]{xy}

\usepackage{float}
\restylefloat{table}

\newcommand{\mc}[1]{\mathcal{#1}}
\newcommand{\mf}[1]{\mathfrak{#1}}
\newcommand{\mb}[1]{\mathbb{#1}}

\newcommand{\ul}[1]{\,\underline {\!#1\!}\,}
\newcommand{\id}{\mathbbm{1}}

\newcommand{\tint}{{\textstyle\int}}

\DeclareMathOperator{\Hom}{Hom}
\DeclareMathOperator{\End}{End}

\DeclareMathOperator{\tr}{tr}

\DeclareMathOperator{\ad}{ad}

\DeclareMathOperator{\Res}{Res}

\DeclareMathOperator{\HR}{HR}

\begin{document}
\mainmatter              
\title{$W$-algebras via Lax type operators}
%
%
\author{Daniele Valeri}
\authorrunning{Daniele Valeri} 
%
\tocauthor{Daniele Valeri}
\institute{School of Mathematics and Statistics, University of Glasgow, G12 8QQ Glasgow, UK\\
\email{daniele.valeri@glasgow.ac.uk}
}
\maketitle              

\begin{abstract}
$W$-algebras are certain algebraic structures associated to a finite dimensional Lie algebra $\mf g$ and a nilpotent element $f$ via
Hamiltonian reduction.
In this note we give a review of a recent approach to the study of (classical affine and quantum finite) $W$-algebras based on the notion of Lax type operators.

For a finite dimensional representation of $\mf g$ a Lax type operator for $W$-algebras is constructed using the 
theory of generalized quasideterminants. This operator carries several pieces of information about the structure and properties of the $W$-algebras
and shows the deep connection of the theory of $W$-algebras with Yangians and integrable Hamiltonian hierarchies of Lax type equations.
%
\end{abstract}
%

\section{Introduction}
\label{sec:intro}

The first \emph{quantum affine $W$-algebra}, the so called Zamolodchikov $W_3$-algebra \cite{Zam85},
appeared in the physics literature in the study of 2-dimensional Conformal Field Theory.
Further generalizations of this algebra were provided soon after \cite{FL88,LF89}.
Physicists thought of these algebras as ``non-linear'' infinite dimensional Lie algebras
extending the Virasoro Lie algebra.
In \cite{FF90} the affine $W$-algebras
$W_\kappa(\mf g,f)$ ($\kappa$ is called the level), for a principal nilpotent element $f\in\mf g$,
were described as vertex algebras obtained via a quantization of the Drinfeld-Sokolov Hamiltonian reduction,
which was used in \cite{DS85} to construct \emph{classical affine $W$-algebras}.
In particular, for $\mf{sl}_2$ one gets the Virasoro vertex algebra, 
and for $\mf{sl}_3$ the Zamolodchikov's $W_3$ algebra.  
The construction was finally generalized to arbitrary nilpotent element $f$
in \cite{KRW03,KW04,KW05}. In these papers, affine $W$-algebras were applied to
representation theory of superconformal algebras.
Quantum affine $W$-algebras may be also considered as an affinization of \emph{quantum finite $W$-algebras} \cite{Pre02} which are a 
natural quantization of Slodowy slices \cite{GG02}.

$W$-algebras are at the cross roads of representation theory and mathematical physics and play important roles
(just to cite some of them)
in applications to integrable systems \cite{DS85,DSKV13}, to Gromov-Witten theory and singularity theory \cite{BM13,ML16},
the geometric Langlands program \cite{FF11,Fre95,FA18,FH16}, four-dimensional gauge theories \cite{AGT10,SV13,BFN16}.

In this note we survey the recent approach to (quantum finite and classical affine) $W$-algebras based on the notion of Lax type 
operators \cite{DSKV16b,DSKV17,DSKVcl,DSKVfin}. For a review of the approach to (classical) $W$-algebras via generators and relations we refer to \cite{DS14}.

Throughout the paper the base field $\mb F$ is a field of characteristic zero.

\smallskip

\noindent{\bf Acknowledgments.} This review is based on the talk I gave at the XIth International Symposium Quantum Theory
and Symmetry in Montr\'eal. I wish to thank the organizers for the invitation and the hospitality.

\section{What is a $W$-algebra?}
\label{sec:what}

$W$-algebras are a rich family of algebraic structures associated to a pair $(\mf g,f)$ consisting of a
finite dimensional reductive 
Lie algebra $\mf g$ and a nilpotent element $f\in\mf g$. They are obtained via Hamiltonian reduction in different categories: Poisson 
algebras, associative algebras and (Poisson) vertex 
algebras.
We should think of them as
algebraic structures underlying
some physical theories with ``extended symmetries''.

\subsection{Fundamental physical theories and corresponding fundamental algebraic structures}
\label{sec:1.1}

In Classical Mechanics the phase space,
describing the possible configurations of a physical system, 
is a Poisson manifold. 
The physical observables are the smooth functions on the manifold,
and they thus form a \emph{Poisson algebra} (PA).

By quantizing this theory we go to Quantum Mechanics.
The observables become non commutative objects,
elements of an \emph{associative algebra} (AA). Hence, the Poisson bracket is replaced by the usual
commutator and the phase space is described as a representation
of this associative algebra.

Going from a finite to an infinite number of degrees of freedom,
we pass from Classical and Quantum Mechanics to
Classical and Quantum Field Theory respectively.
The algebraic structure corresponding to an arbitrary Quantum Field Theory
is still to be understood,
but in the special case of chiral quantum fields
of a 2-dimensional Conformal Field Theory (CFT) the adequate algebraic structure is a \emph{vertex algebra} (VA)
\cite{Bor86},
and its quasi-classical limit is known as \emph{Poisson vertex algebra} (PVA) \cite{DSK06}.

Hence, the algebraic counterparts of the four fundamental frameworks of physical theories
can be put in the following diagram:
%
\begin{equation}\label{maxi}
\UseTips
\xymatrix{
PVA\,\,\,\,\,\,  \ar[d]^{\text{Zhu}}
\ar@/^1pc/@{.>}[r]^{\text{quantization}} &
\ar[l]^{\text{cl.limit}}
\,\,\,\,\,\,VA 
\ar[d]_{\text{Zhu}} \\
PA\,\,\,\,\,\,  \ar@/^1pc/@{.>}[u]^{\text{affiniz.}}
\ar@/_1pc/@{.>}[r]_{\text{quantization}} &
\ar[l]_{\text{cl.limit}}
\,\,\,\,\,\, AA \ar@/_1pc/@{.>}[u]_{\text{affiniz.}}
}
\end{equation}
%
The straight arrows in the above diagram correspond to canonical functors and have the following meaning.
Given a filtered AA (respectively VA),
its associated graded algebra is a PA (respectively PVA) called its \emph{classical limit}.
Moreover, starting from a positive energy VA 
(respectively PVA)
we can construct an AA (resp. PA)
governing its representation theory, known as its \emph{Zhu algebra} \cite{Zhu96}.
The processes of going from a classical theory  to a quantum theory (``quantization''),
or from finitely many to infinitely many degrees of freedom (``affinization'') are not functorial
and they are thus represented with dotted arrows.

\subsubsection{(Poisson) vertex algebras.}
PVAs provide a convenient framework to study Hamiltonian partial differential equations. Recall from \cite{BDSK09} that a PVA
is a differential algebra,  i.e. a unital commutative associative algebra with a derivation $\partial$,
endowed with a $\lambda$-\emph{bracket},
i.e. a bilinear (over $\mb F$) map $\{\cdot\,_\lambda\,\cdot\}:\,\mc V\times\mc V\to\mc V[\lambda]$, 
satisfying the following axioms ($a,b,c\in\mc V$):
\begin{enumerate}[(i)]
\item
sesquilinearity:
$\{\partial a_\lambda b\}=-\lambda\{a_\lambda b\}$,
$\{a_\lambda\partial b\}=(\lambda+\partial)\{a_\lambda b\}$;
\item
skewsymmetry:
$\{b_\lambda a\}=-\{a_{-\lambda-\partial} b\}$;
\item
Jacobi identity:
$\{a_\lambda \{b_\mu c\}\}-\{b_\mu\{a_\lambda c\}\}
=\{\{a_\lambda b\}_{\lambda+\mu}c\}$;
\item
(left) Leibniz rule:
$\{a_\lambda bc\}=\{a_\lambda b\}c+\{a_\lambda c\}b$.
\end{enumerate}
Applying skewsymmetry to the left Leibniz rule
we get
\begin{enumerate}[(i)]
\setcounter{enumi}{4}
\item
right Leibniz rule:
$\{ab_\lambda c\}=\{a_{\lambda+\partial} c\}_\to b+\{b_{\lambda+\partial} c\}_\to a$.
\end{enumerate}
In (ii) and (iv) we use the following notation: if
$\{a_\lambda b\}=\sum_{n\in\mb Z_+}\lambda^n \alpha_n\in\mc V[\lambda]$,
then
$\{a_{\lambda+\partial}b\}_{\rightarrow}c
=\sum_{n\in\mb Z_+}\alpha_n(\lambda+\partial)^nc\in\mc V[\lambda]$
and
$\{a_{-\lambda-\partial}b\}=\sum_{n\in\mb Z_+}(-\lambda-\partial)^n\alpha_n\in\mc V[\lambda]$
(if there is no arrow we move $\partial$ to the left).

We denote by $\tint:\,\mc V\to\mc V/\partial\mc V$ the canonical quotient map
of vector spaces.
Recall that, if $\mc V$ is a PVA,
then $\mc V/\partial\mc V$ carries a well defined Lie algebra structure given by
$\{\tint f,\tint g\}=\tint\{f_\lambda g\}|_{\lambda=0}$,
and we have a representation of the Lie algebra $\mc V/\partial\mc V$ on $\mc V$
given by $\{\tint f,g\}=\{f_\lambda g\}|_{\lambda=0}$.
A \emph{Hamiltonian equation} on $\mc V$ associated to a \emph{Hamiltonian functional} 
$\tint h\in\mc V/\partial\mc V$ is the evolution equation 
\begin{equation}\label{ham-eq}
\frac{du}{dt}=\{\tint h,u\}\,\,, \,\,\,\, u\in\mc V\,.
\end{equation}
The minimal requirement for \emph{integrability} is to have an infinite collection
of linearly independent integrals of motion in involution:
$$
\tint h_0=\tint h,\,\tint h_1,\,\tint h_2,\,\dots\,
\,\text{ s.t. }\,\, \{\tint h_m,\tint h_n\}=0\,\,\text{ for all }\,\, m,n\in\mb Z_{\geq0}
\,.
$$
In this case, we have the \emph{integrable hierarchy} of Hamiltonian equations
$$
\frac{du}{dt_n}=\{\tint h_n,u\}\,\,, \,\,\,\, n\in\mb Z_{\geq0}
\,.
$$

\begin{example}\label{exa:vir}
The Virasoro-Magri PVA on the algebra of differential polynomials $\mc V=\mb C[u,u',u'',\dots]$ is defined by letting
$$
\{u_\lambda u\}=(2\lambda+\partial)u+\lambda^3\,,
$$
and extending it to a $\lambda$-bracket for the whole $\mc V$ using sesquilinearity and Leibniz rules.
Let $\tint h=\tint\frac{u^2}{2}$. Then the corresponding Hamiltonian equation \eqref{ham-eq}
is the famous \emph{KdV equation}:
$$
\frac{du}{dt}=u'''+3uu'
\,.
$$
Using the Lenard-Magri scheme of integrability \cite{Mag78} it can be shown that it belongs to an integrable hierarchy.
\end{example}

\subsubsection*{Vertex algebras.}
VAs were introduced in \cite{Bor86}. Following \cite{DSK06}, we provide here a ``Poisson-like'' definition
using $\lambda$-brackets.
A VA is a 
(not necessarily commutative nor associative) unital algebra
$V$ with a derivation $\partial$
endowed with
a $\lambda$-bracket
$[\cdot_{\lambda}\cdot]:V\times V\longrightarrow V [\lambda]$
satisfying sesquilinearity, skewsymmetry, Jacobi identity, and, moreover ($a,b,c\in V$):
\begin{enumerate}[(i)]
\item
quasicommutativity:
$ab-ba=\int_{-\partial}^0[a_\lambda b]d\lambda$;
\item
quasiassociativity:
$(ab)c-a(bc)=(|_{\lambda=\partial}a)\int_0^\lambda[b_\mu c]d\mu +(|_{\lambda=\partial}b)\int_0^\lambda[a_\mu c]d\mu$;
\item
noncommutative Wick formula:
$[a_\lambda bc]=[a_\lambda b]c+b[a_\lambda c]+\int_0^\lambda[[a_\lambda b]_{\mu}c]d\mu$.
\end{enumerate}
We refer to \cite{DSK06} for explanations about the notation.
As before, we denote by $\tint:\,V\to V/\partial V$ the canonical quotient map
of vector spaces. If $V$ is a VA,
then $V/\partial V$ carries a well defined Lie algebra structure given by
$[\tint f,\tint g]=\tint[f_\lambda g]|_{\lambda=0}$,
and we have a representation of the Lie algebra $V/\partial V$ on $V$
given by $[\tint f,g]=[f_\lambda g]|_{\lambda=0}$.
A quantum integrable system consists in a collection of infinitely many linearly independent elements $\tint h_m\in V/\partial V$,
$m\in\mb Z_{\geq0}$, in involution. 

\begin{example}
A VA is commutative if $[a_\lambda b]=0$, for every $a,b\in V$. It follows immediately from the definition that the category of commutative VAs is the same as the category of differential algebras.
\end{example}

\begin{remark}
The (not necessarily commutative nor associative) product in a VA corresponds to the \emph{normally ordered product} of quantum fields in a CFT,
while the $\lambda$-bracket encodes the
singular part of their \emph{operator product expansion} (OPE).  We give a naive explanation of the latter sentence in a 
particular case. Consider the VA $\lambda$-bracket of a Virasoro element  $u$ (recall Example \ref{exa:vir} for its PVA analogue)
$$
[u_\lambda u]=(2\lambda +\partial)u+\frac{c}{12}\lambda^3
\,,
$$
where $c\in\mb C$ is called the central charge. Replace, in the above relation, $u$ by a quantum field, say $T(w)$, $\partial$ by
$\partial_w$ and $\lambda$ by $\partial_w$ acting on the rational function $\frac{1}{z-w}$. Then we get
$$
[T(w)_{\partial_w} T(w)]_\to\frac{1}{z-w}
=\frac{\partial_wT(w)}{z-w}+\frac{2T(w)}{(z-w)^2}+\frac{c/2}{(z-w)^4}
\,,
$$
which is the singular part of the OPE of the stress-energy tensor in CFT.

\end{remark}

\subsection{A toy model}
\label{sec:1.2}

The simplest example when all four objects in diagram \eqref{maxi} can be constructed,
is obtained starting with a  finite-dimensional Lie algebra
$\mf g$,
with Lie bracket $[\cdot\,,\,\cdot]$,
and with a non-degenerate invariant symmetric bilinear form $(\cdot\,|\,\cdot)$.

The \emph{universal enveloping algebra} of $\mf g$,
usually denoted by $U(\mf g)$, is an associative algebra,
and its classical limit is the symmetric algebra $S(\mf g)$,
with the Kirillov-Kostant Poisson bracket.

Furthermore, we have also a Lie conformal algebra
$\text{Cur}\,\mf g=(\mb F[\partial]\otimes\mf g)\oplus\mb FK$,
with the following $\lambda$-bracket:
\begin{equation}\label{20140401:eq2}
[a_\lambda b]
=
[a,b]+(a|b)K\lambda
\,\,,\,\,\,\,
[a_\lambda K]=
0
\,\,,\,\,\,\,
\text{ for }\,\, a,b\in\mf g
\,.
\end{equation}
The universal enveloping vertex algebra of $\text{Cur}\,\mf g$
is the so-called \emph{universal affine vertex algebra} $V(\mf g)$,
and its classical limit is the algebra of differential polynomials
$\mc V(\mf g)=S(\mb F[\partial]\mf g)$,
with the PVA $\lambda$-bracket defined by \eqref{20140401:eq2}. We refer to \cite{DSK06}
for the definition of the latter structures and the construction of the corresponding Zhu maps.
Thus, 
we get the following basic example of diagram \eqref{maxi}:
%
\begin{equation}\label{maxi2}
\UseTips
\xymatrix{
\mc V(\mf g)\,\,\,\,\,\,  \ar[d]^{\text{Zhu}}
\ar@/^1pc/@{.>}[r]^{\text{quantization}} &
\ar[l]^{\text{cl.limit}}
\,\,\,\,\,\,V(\mf g) 
\ar[d]_{\text{Zhu}} \\
S(\mf g)\,\,\,\,\,\,  \ar@/^1pc/@{.>}[u]^{\text{affiniz.}}
\ar@/_1pc/@{.>}[r]_{\text{quantization}} &
\ar[l]_{\text{cl.limit}}
\,\,\,\,\,\, U(\mf g) \ar@/_1pc/@{.>}[u]_{\text{affiniz.}}
}
\end{equation}

\subsection{Hamiltonian reduction}
\label{sec:1.3}

All the four algebraic structures in diagram \eqref{maxi}
admit a Hamiltonian reduction. We review here only the case for associative algebras.
Recall that the Hamiltonian reduction of a unital associative algebra $A$
by a pair $(B,I)$,
where $B\subset A$ is a unital associative subalgebra and 
$I\subset B$ is a two sided ideal,
is the following unital associative algebra:
\begin{equation}\label{20140401:eq4}
W
=
W(A,B,I)
=
\big(A\big/A I\big)^{\ad B}
\,.
\end{equation}
where $\ad B$ denotes the usual adjoint action given by the commutator in an associative algebra (note that $B$ acts on
$A/AI$ both by left and right multiplication). 
It is not hard to show that the obvious associative product on $W$
is well-defined.

Now, let $\{e,2x,f\}\subset\mf g$ be an $\mf{sl}_2$-triple, and let
\begin{equation}\label{eq:dec}
\mf g=\bigoplus_{\substack{j=-d\\j\in\frac12\mb Z}}^d\mf g_j
\,,
\end{equation}
be the $\ad x$-eigenspace decomposition. We can perform the Hamiltonian reduction of $A=U(\mf g)$ as follows.
Let $B=U(\mf g_{>0})$ and $I\subset B$ be the two sided ideal generated by the set
\begin{equation}\label{set}
\big\{m-(f|m)\,\big|\,m\in\mf g_{\geq1}\big\}
\,.
\end{equation}
Applying the Hamiltonian reduction \eqref{20140401:eq4} with the above data we get the so-called \emph{quantum finite $W$-algebra}
(it first appeared in \cite{Pre02})
$$
W^{\text{fin}}(\mf g,f)=\big(U(\mf g)\big/U(\mf g)\{m-(f|m)\,\big|\,m\in\mf g_{\geq1}\}\big)^{\ad \mf g_{>0}}
\,.
$$

The Hamiltonian reduction \eqref{20140401:eq4} still makes sense if we replace associative algebras with
PVAs (respectively, PAs), and we can perform it with $A=\mc V(\mf g)$, $B=\mc V(\mf g_{>0})$ and
$I\subset B$ the differential algebra ideal generated by the set \eqref{set} (respectively, $A=S(\mf g)$, $B=S(\mf g_{>0})$ and
$I\subset B$ the ideal generated by the set \eqref{set}).
As a result we get the so-called \emph{classical affine} $W$-algebra $\mc W^{\text{aff}}(\mf g,f)$ (respectively, \emph{classical finite}
$W$-algebra $\mc W^{\text{aff}}(\mf g,f)$), see \cite{DSKV16a} for further details.  

Unfortunately,
a similar construction of a Hamiltonian reduction for vertex algebras is not known, and 
the \emph{quantum affine $W$-algebra} $W^{\text{aff}}(\mf g,f)$ 
is constructed using a cohomological approach \cite{FF90,KW04}.

\subsection{From the toy model to $W$-algebras}

Let $\mf g$ be a finite dimensional reductive Lie algebra, and let $f\in\mf g$ be a nilpotent element. By the Jacobson-Morozov Theorem it can be
embedded in an $\mf{sl}_2$-triple $\{e,2x,f\}\subset\mf g$.
Applying the machinery described in Section \ref{sec:1.3} we thus obtain a Hamiltonian reduction of the whole diagram \eqref{maxi2}:
\begin{equation}\label{maxi3}
\UseTips
\xymatrix
{
\mc V(\mf g)\ar[dd]_{\text{Zhu}}\ar[dr]^{\HR_f}
&
&
\ar[ll]_{\text{cl.limit}}
V(\mf g) \ar[dd]|!{[d];[d]}\hole
\ar[dr]^{\HR_f}
&
\\
&
\mc W^{\text{aff}}(\mf g,f)
\ar[dd]_(.35){\text{Zhu}}
&
&
W^{\text{aff}}(\mf g,f)
\ar[ll]_(.62){\text{cl.limit}}
\ar[dd]_(.35){\text{Zhu}}
\\
S(\mf g)
\ar[dr]^{\HR_f}
&
&
U(\mf g)
\ar[ll]|!{[l];[l]}\hole
\ar[dr]^{\HR_f}
&
\\
&
\mc W^{\text{fin}}(\mf g,f)
&
&
W^{\text{fin}}(\mf g,f)
\ar[ll]_{\text{cl.limit}}
}
\end{equation}
It is a convention to use the calligraphic $\mc W$ to denote objects appearing in the ``classical'' column of diagram \eqref{maxi3}
and the block letter $W$ to denote objects appearing in the ``quantum'' column of the same diagram.
Also the upper label ``fin'' (resp. ``aff'') is used to denote objects appearing in the ``finite'' (resp. ``affine'') row of  diagram \eqref{maxi3}, corresponding to physical theories with a finite (resp. infinite) number of degrees of freedom.

Hence, as we can see from diagram \eqref{maxi3}, $W$-\emph{algebras} provide a very rich family of examples
which appear in all the four fundamental aspects in diagram \eqref{maxi}.
Each of these classes of algebras
was introduced and studied separately, with different applications in mind.
The relations between them became fully clear later, see \cite{GG02,DSK06,DSKV16a} for further details.

\subsubsection{Classical finite $W$-algebras.}
The classical finite $W$-algebra $\mc W^{\text{fin}}(\mf g,f)$ 
is a PA,
which can be viewed as the algebra of functions 
on the so-called \emph{Slodowy slice} $\mc S(\mf g,f)$,
introduced by Slodowy
while studying the singularities 
associated to the coadjoint nilpotent orbits of $\mf g$ \cite{Slo80}.

\subsubsection{Finite $W$-algebras.}
The first appearance of the finite $W$-algebras $W^{\text{fin}}(\mf g,f)$ 
was in a paper of Kostant \cite{Kos78}.
He constructed the finite $W$-algebra for principal nilpotent $f\in\mf g$
(in which case it is commutative),
and proved that it is isomorphic to the center 
of the universal enveloping algebra $U(\mf g)$.
The construction was then extended in \cite{Lyn79}
for even nilpotent element $f\in\mf g$.
The general definition of finite $W$-algebras $W^{\text{fin}}(\mf g,f)$,
for an arbitrary nilpotent element $f\in\mf g$,
appeared later in a paper by Premet \cite{Pre02}.
Finite $W$-algebras have deep connection with geometry and representation theory of simple
finite-dimensional Lie algebras,
with the theory of primitive ideals, and the Yangians, see \cite{Mat90,Pre02,Pre05,BrK06}.

\subsubsection{Classical affine $W$-algebras.}
The classical affine $W$-algebras $\mc W^{\text{aff}}(\mf g,f)$ 
were introduced, for principal nilpotent element $f$,
in the seminal paper of Drinfeld and Sokolov \cite{DS85}.
They were introduced as Poisson algebras of functions 
on an infinite dimensional Poisson manifold,
and they were used to study KdV-type integrable bi-Hamiltonian hierarchies of PDE's,
nowadays known as Drinfeld-Sokolov hierarchies.
Later, there have been several papers aimed at the construction of generalized Drinfeld-Sokolov hierarchies, \cite{dGHM92,FHM93,BdGHM93,DF95,FGMS95,FGMS96}.
In \cite{DSKV13},
the classical $W$-algebras $\mc W^{\text{aff}}(\mf g,f)$
were described as PVA,
and the theory of generalized Drinfeld-Sokolov hierarchies
was formalized in a more rigorous and complete way \cite{DSKV14a,DSKV16b,DSKVcl}.

\subsubsection{Quantum affine $W$-algebras.}
They have been extensively discussed in the Introduction.
A review
of the subject up to the early 90's may be found in the
collection of a large number of reprints on $W$-algebras \cite{BS95}.
Recently, it has been shown that they are at the base of an unexpected connections of vertex algebras with the geometric
invariants called the Higgs branches in the four dimensional $N=2$ superconformal field theories \cite{BR17,Ara17}.

\section{Linear algebra intermezzo}
\label{sec:linear}

\subsection{Set up}\label{sec:setup}
Let $\mf g$ be a finite dimensional reductive Lie algebra,
let $\{f,2x,e\}\subset\mf g$ be an $\mf{sl}_2$-triple
and let \eqref{eq:dec}
be the corresponding $\ad x$-eigenspace decomposition.
In Sections \ref{sec:finite} and \ref{sec:affine} we will use the projection map
$\pi_{\leq\frac12}:\mf g\to \mf g_{\leq \frac12}=\oplus_{k\leq\frac12}\mf g_k$ with kernel
$\mf g_{>\frac12}=\oplus_{k>\frac12}\mf g_k$.

Let $\varphi:\,\mf g\to\End V$ be a faithful representation of $\mf g$
on an $N$-dimensional vector space $V$.
Throughout the paper we shall often use the following convention:
we denote by lowercase Latin letters elements of the Lie algebra $\mf g$,
and by the same uppercase letters the corresponding (via $\varphi$)
elements of $\End V$.
For example, $F=\varphi(f)$ is a nilpotent endomorphism of $V$.
Moreover, $X=\varphi(x)$ is a semisimple endomorphism of $V$ with half-integer eigenvalues.
The corresponding $X$-eigenspace decomposition of $V$ is
\begin{equation}\label{eq:grading_V}
V=\bigoplus_{k\in\frac12\mb Z}V[k]
\,.
\end{equation}
Note that $\frac d2$ is the largest $X$-eigenvalue in $V$.

Recall that the trace form on $\mf g$ associated to the representation $V$ 
is, by definition,
\begin{equation}\label{20170317:eq1}
(a|b)=\tr_V(AB)\,,
\qquad
a,b\in\mf g
\,,
\end{equation}
and we assume that it is non-degenerate.
Let $\{u_i\}_{i\in I}$ be a basis of $\mf g$ 
compatible with the $\ad x$-eigenspace decomposition \eqref{eq:dec},
i.e. $I=\sqcup_k I_k$ where $\{u_i\}_{i\in I_k}$ is a basis of $\mf g_k$.
We also denote $I_{\leq\frac12}=\sqcup_{k\leq\frac12}I_k$.
Moreover, we shall also need, in Section \ref{sec:affine}, 
that $\{u_i\}_{i\in I}$ contains a basis $\{u_i\}_{i\in I_f}$ of $\mf g^f=\{a\in\mf g\mid [a,f]=0\}$, the centraliser of $f$ in $\mf g$.
Let $\{u^i\}_{i\in I}$ be the basis of $\mf g$ dual to $\{u_i\}_{i\in I}$ with respect 
to the form \eqref{20170317:eq1},
i.e. $(u_i|u^j)=\delta_{i,j}$.
According to our convention,
we denote by $U_i=\varphi(u_i)$ and $U^i=\varphi(u^i)$, $i\in I$, 
the corresponding endomorphisms of $V$.

In Sections \ref{sec:finite} and \ref{sec:affine} we will consider the following important element
\begin{equation}\label{20170623:eq4}
U=\sum_{i\in I}u_i U^i\in\mf g\otimes\End V
\,.
\end{equation}
Here and further we are omitting the tensor product sign.

Furthermore, the following endomorphism of $V$, which we will call the \emph{shift matrix},
will play an important role in Section \ref{sec:finite}
\begin{equation}\label{eq:D}
D
=
-\sum_{i\in I_{\geq1}}U^iU_i
\,\in\End V
\,.
\end{equation}
Finally,
we denote by $\Omega_V\in\End V\otimes\End V$
the permutation map:
\begin{equation}\label{Omega}
\Omega_V(v_1\otimes v_2)=v_2\otimes v_1
\,\,\text{ for all }\,\,v_1,v_2\in V
\,.
\end{equation}
Using Sweedler's notation we write $\Omega_V=\Omega_V^\prime\otimes\Omega_V^{\prime\prime}$
to denote, as usual, a sum of monomials in $\End V\otimes\End V$.
Suppose that $V$ has a non-degenerate bilinear form $\langle\cdot\,|\,\cdot\rangle:\,V\times V\to\mb F$,
which is symmetric or skewsymmetric:
\begin{equation}\label{eq:epsilon}
\langle v_1|v_2\rangle=\epsilon\langle v_2|v_1\rangle
\,,\,\,v_1,v_2\in V
\,,\,\,\text{ where }
\epsilon\in\{\pm1\}
\,.
\end{equation}
Then, we denote by
\begin{equation}\label{Omega-dagger}
\Omega_V^\dagger=(\Omega_V')^\dagger\otimes\Omega_V''
\,,
\end{equation}
where
$A^\dagger$ is the adjoint of $A\in\End V$ with respect to \eqref{eq:epsilon}.

\subsection{
The ``identity'' notation
}
\label{sec:8a.2.5}

Let $U\subset V$ be a subspace of $V$,
and assume that there is ``natural'' splitting $V=U\oplus U^\prime$.
We shall denote, with an abuse of notation,
by $\id_U$ both the identity map $U\stackrel{\sim}{\longrightarrow}U$,
the inclusion map $U\hookrightarrow V$,
and the projection map $V\twoheadrightarrow U$ with kernel $U^\prime$.
The correct meaning of $\id_U$ should be clear from the context.

\subsection{
Generalized quasi-determinants
}
\label{sec:8a.3}

Let $R$ be a unital associative algebra and let $V$ be a finite dimensional vector space
with direct sum decompositions $V=U\oplus U^\prime=W\oplus W^\prime$.
Assume that $A\in R\otimes \End(V)$ and 
$\id_WA^{-1}\id_U\in R\otimes\Hom(U,W)$ are invertible.
The (generalized) \emph{quasideterminant} of $A$ with respect to $U$ and $W$,
cf. \cite{GGRW05,DSKV16b}, is defined as
\begin{equation}\label{eq:quasidet}
|A|_{U,W}
:=
(\id_{W} A^{-1}\id_{U})^{-1}\in R\otimes\Hom(W,U)
\,.
\end{equation}
\begin{remark}
Provided that both $A$ and $\id_{U'}A\id_{W'}$ are invertible, it is possible to write the
generalised quasideterminant \eqref{eq:quasidet} in the more explicit form $|A|_{U,W}
=
\id_U A\id_W
-
\id_U A\id_{W^\prime}
(\id_{U^\prime} A\id_{W^\prime})^{-1}
\id_{U^\prime} A\id_W$.
\end{remark}

\section{Quantum finite $W$-algebras and (twisted) Yangians}
\label{sec:finite}

\subsection{Lax type operators for quantum finite $W$-algebras}
We introduce some important $\End V$-valued polynomials in $z$,
and Laurent series in $z^{-1}$, with coefficients in $U(\mf g)$.
The first one is (cf. \eqref{20170623:eq4})
\begin{equation}\label{eq:A}
A(z)=z\id_V+U=z\id_V+\sum_{i\in I}u_i U^i
\,\,
\in U(\mf g)[z]\otimes\End(V)
\,.
\end{equation}
(As in Section \ref{sec:linear}, we are dropping the tensor product sign.)
Another important operator is
(keeping the same notation as in \cite{DSKV17})
\begin{equation}\label{eq:Arho}
A^{\rho}(z)
=
z\id_V+F+\pi_{\leq\frac12}U
=
z\id_V+F+\sum_{i\in I_{\leq\frac12}}u_i U^i
\,\,\in U(\mf g)[z]\otimes\End V
\,.
\end{equation}

Now we introduce 
the Lax operator $L(z)$.
Consider the generalized quasideterminant (cf. \eqref{eq:quasidet})
\begin{equation}\label{eq:tildeL}
\widetilde L(z)
=
|A^\rho(z)+D|_{V[\frac d2],V[-\frac d2]}
=
\Big(\id_{V[-\frac d2]}\big(z\id_V+F+\pi_{\leq\frac12}U+D\big)^{-1}\id_{V[\frac d2]}\Big)^{-1}
\,,
\end{equation}
where $\id_{V[-\frac d2]}$ and $\id_{V[\frac d2]}$ are defined in Section \ref{sec:8a.2.5}
(using the obvious splittings of $V$ given by the grading \eqref{eq:grading_V}),
$A^{\rho}(z)$ is defined in equation \eqref{eq:Arho}
and $D$ is the ``shift matrix'' \eqref{eq:D}.

Let us denote by $\bar1$ the image of $1\in U(\mf g)$ in the quotient 
$U(\mf g)\big/U(\mf g)\{m-(f|m)\,\big|\,m\in\mf g_{\geq1}\}$.
The Lax operator $L(z)$ is defined as the image of $\widetilde L(z)$ in this quotient:
\begin{equation}\label{eq:L}
L(z)
=
L_{\mf g,f,V}(z)
:=
\widetilde L(z)\bar 1
\,.
\end{equation}
The first main result in \cite{DSKVfin} can be summarized as follows.

\begin{theorem}\label{thm:L1}
\begin{enumerate}[(a)]
\item
The operator $A^\rho(z)+D$ is invertible in
$U(\mf g)((z^{-1}))\otimes\End V$, and the 
operator $\id_{V[-\frac d2]}(A^\rho(z)+D)^{-1}\id_{V[\frac d2]}$
is invertible in
$U(\mf g)((z^{-1}))\otimes\Hom\big(V\big[-\frac d2\big],V\big[\frac d2\big]\big)$.
Hence,
the quasideterminant defining $\widetilde L(z)$ (cf. \eqref{eq:tildeL}) exists
and lies in $U(\mf g)((z^{-1}))\otimes\Hom\big(V\big[-\frac d2\big],V\big[\frac d2\big]\big)$.
\item
The entries of the coefficients of the operator $L(z)$ defined in \eqref{eq:L} 
lie in the $W$-algebra $W(\mf g,f)$:
$$
L(z):=|z\id_V+F+\pi_{\leq\frac12}U+D|_{V[\frac d2],V[-\frac d2]}\bar 1
\,\in
W(\mf g,f)((z^{-1}))\otimes\Hom\big(V\big[-\frac d2\big],V\big[\frac d2\big]\big)
\,.
$$
\end{enumerate}
\end{theorem}

\begin{remark}
For $\mf g=\mf{gl}_N$ and $V=\mb F^N$ the standard representation, equation \eqref{eq:tildeL}
may be used to find a generating set (in the sense of PBW Theorem) for the quantum finite $W$-algebra,
see \cite{laura} for more details.
\end{remark}

\subsection{The generalized Yangian identity}

Let $\alpha,\beta,\gamma\in\mb F$.
Let $R$ be a unital associative algebra,
and let $V$ be an $N$-dimensional vector space.
For $\beta\neq0$,
we also assume, as in Section \ref{sec:setup},
that $V$ is endowed with a non-degenerate
bilinear form $\langle\cdot\,|\,\cdot\rangle:\,V\times V\to\mb F$
which we assume to be symmetric or skewsymmetric,
and we let $\epsilon=+1$ and $-1$ respectively.
Again, when denoting an element of $R\otimes\End(V)$
or of  $R\otimes\End(V)\otimes\End(V)$,
we omit the tensor product sign on the first factor,
i.e. we treat elements of $R$ as scalars.

The \emph{generalized} $(\alpha,\beta,\gamma)$-\emph{Yangian identity}
for $A(z)\in R((z^{-1}))\otimes\End(V)$
is the following identity, holding in $R[[z^{-1},w^{-1}]][z,w]\otimes\End(V)\otimes\End(V)$:
\begin{equation}\label{eq:gener-yangV}
\begin{array}{l}
\displaystyle{
\vphantom{\Big(}
(z-w+\alpha\Omega_V)
(A(z)\otimes\id_V)
(z+w+\gamma-\beta\Omega_V^\dagger)
(\id_V\otimes A(w))
} \\
\displaystyle{
\vphantom{\Big(}
=
(\id_V\otimes A(w))
(z+w+\gamma-\beta\Omega_V^\dagger)
(A(z)\otimes\id_V)
(z-w+\alpha\Omega_V)
\,.}
\end{array}
\end{equation}
Recall that $\Omega_V$ and $\Omega_V^\dagger$ are defined by equations \eqref{Omega} and
\eqref{Omega-dagger} respectively.
\begin{remark}\label{20170704:rem2}
In the special case $\alpha=1$, $\beta=\gamma=0$,
equation \eqref{eq:gener-yangV} coincides with the so called RTT presentation
of the Yangian of $\mf{gl}(V)$, cf. \cite{Mol07,DSKV17}. 
Moreover,
in the special case $\alpha=\beta=\frac12$, $\gamma=0$,
equation \eqref{eq:gener-yangV} coincides with the so called RSRS presentation
of the extended twisted Yangian of $\mf g=\mf{so}(V)$ or $\mf{sp}(V)$, 
depending on whether $\epsilon=+1$ or $-1$, cf. \cite{Mol07}. 
Hence, if $A(z)\in R((z^{-1}))\otimes\End V$
satisfies the generalized $(\frac12,\frac12,0)$-Yangian identity 
we automatically have an algebra homomorphism from 
the extended twisted Yangian $X(\mf g)$ to the algebra $R$.
If, moreover, $A(z)$ satisfies the symmetry condition
(required in the definition of twisted Yangian in \cite{Mol07})
$$
A^\dagger(-z)-\epsilon A(z)
=
-\frac{A(z)-A(-z)}{4z}
\,,
$$
then we have an algebra homomorphism from 
the twisted Yangian $Y(\mf g)$ to the algebra $R$.
\end{remark}

\subsection{Quantum finite $W$-algebras and (extended) twisted Yangians}

Let $\mf g$ be one of the classical Lie algebras $\mf{gl}_N$, $\mf{sl}_N$, $\mf{so}_N$ or $\mf{sp}_N$,
and let $V=\mb F^N$ be its standard representation
(endowed, in the cases of $\mf{so}_N$ and $\mf{sp}_N$,
with a non-degenerate symmetric or skewsymmetric bilinear form, respectively).
Then, the operator
$A(z)$ defined in equation \eqref{eq:A}
satisfies the generalized Yangian identity \eqref{eq:gener-yangV},
where $\alpha,\beta,\gamma$ are given by the following table:
$$
\begin{tabular}{c|lll}
\vphantom{\Big(}
$\mf g$ & \,\,$\alpha$\,\, & \,\,$\beta$\,\, & \,\,$\gamma$\,\, \\
\hline 
\vphantom{\Big(}
$\mf{gl}_N$ or $\mf{sl}_N$ &  $1$ & $0$ & $0$ \\
\vphantom{\Big(}
$\mf{so}_N$ or $\mf{sp}_N$ &  $\frac12$ & $\frac12$ & $\frac\epsilon2$ \\
\end{tabular}
$$
Note that $V[\frac{d}{2}]\cong V[-\frac{d}{2}]$.
Fix and isomorphism $\chi:V[\frac{d}{2}]\stackrel{\cong}{\longrightarrow}V[-\frac{d}{2}]$.
Then, $\chi\circ L(z)\in W(\mf g,f)((z^{-1}))\otimes \End(V[-\frac d2])$. By an abuse of notation,
we still denote this operator by $L(z)$. We also let $n=\dim V[-\frac d2]$.

The second main result in \cite{DSKVfin} states that, for classical Lie algebras, the Lax operator defined in \eqref{eq:L} also satisfies a  generalized Yangian identity.
\begin{theorem}\label{thm:main2}
The operator $L(z)\in W(\mf g,f)((z^{-1}))\otimes\End(V[-\frac d2])$
defined by \eqref{eq:tildeL}-\eqref{eq:L}
(cf. Theorem \ref{thm:L1}) satisfies the generalized Yangian identity \eqref{eq:gener-yangV}
with the values of $\alpha,\beta,\gamma$ as in the following table:
$$
\begin{tabular}{c|lll}
\vphantom{\Big(}
$\mf g$ & \,\,$\alpha$\,\, & \,\,$\beta$\,\, & \,\,$\gamma$\,\, \\
\hline 
\vphantom{\Big(}
$\mf{gl}_N$ or $\mf{sl}_N$ &  $1$ & $0$ & $0$ \\
\vphantom{\Big(}
$\mf{so}_N$ or $\mf{sp}_N$  & $\frac12$ & $\frac12$ & $\frac{\epsilon-N+n}2$ \\
\end{tabular}
$$
\end{theorem}
By Theorem \ref{thm:main2} and Remark \ref{20170704:rem2} we have an algebra homomorphism from the extended twisted Yangian
$X(\bar{\mf g})$ ($\bar{\mf g}$ depends on the pair $(\mf g,f)$) to the quantum finite $W$-algebra
$W(\mf g,f)$. A stronger result has been obtained for $\mf g=\mf{gl}_N$ by Brundan and Kleshchev in 
\cite{BrK06} where quantum finite $W$-algebras were constructed as truncated shifted Yangians
(which are subquotients of the Yangian for $\mf{gl}_N$).

\section{Classical affine $W$-algebras and integrable hierarchies of Lax type equations}
\label{sec:affine}

\subsection{Lax type operators for classical affine $W$-algebras}
For classical affine $W$-algebras the discussion is similar to the one in Section \ref{sec:finite} but in a different setting: we need to substitute
polynomials and Laurent series with differential operators and pseudodifferential operators respectively
(see \cite{DSKVcl} for a review of their basic properties).

Consider the differential operators
$$
A(\partial)=\partial\id_V+U=\partial\id_V+\sum_{i\in I}u_i U^i
\,\,
\in \mc V(\mf g)[\partial]\otimes\End(V)
$$
and
$$
A^{\rho}(\partial)
=
\partial\id_V+F+\pi_{\leq\frac12}U
=
\partial\id_V+F+\sum_{i\in I_{\leq\frac12}}u_i U^i
\,\,\in \mc V(\mf g_{\leq\frac12})[\partial]\otimes\End V
\,.
$$
Recall from \cite{DSKV13} that in the classical affine case we have $\mc W(\mf g,f)\subset\mc V(\mf g_{\leq\frac12})$ and that there exists a differential algebra isomorphism
$w:\mc V(\mf g^f)\stackrel{\sim}{\longrightarrow}\mc W(\mf g,f)$.
Consider the generalized quasideterminant (cf. \eqref{eq:quasidet})
\begin{equation}\label{eq:Laff}
L(\partial)
=
|A^\rho(\partial)|_{V[\frac d2],V[-\frac d2]}
=
\Big(\id_{V[-\frac d2]}\big(\partial\id_V+F+\pi_{\leq\frac12}U\big)^{-1}\id_{V[\frac d2]}\Big)^{-1}
\,.
\end{equation}
The following result has been proved in \cite{DSKVcl}.
\begin{theorem}\label{thm:main}
$L(\partial)\in\mc W(\mf g,f)((\partial^{-1}))\otimes\Hom\big(V\big[-\frac d2\big],V\big[\frac d2\big]\big)
$ and 
\begin{equation}\label{eq:Lw}
L(\partial)
=
\Big(\id_{V[-\frac d2]}\big(\partial\id_V+F+\sum_{i\in I_{f}}w(u_i) U^i\big)^{-1}\id_{V[\frac d2]}\Big)^{-1}
\,.
\end{equation}
\end{theorem}
The above theorem consists of two statements.
First, it claims that $L(\partial)$ is well defined, i.e. both inverses in formula
\eqref{eq:Laff} can be carried out in the algebra
of pseudodifferential operators with coefficients in $\mc V(\mf g_{\leq\frac12})$,
and that the coefficients of $L(\partial)$
lie in the $\mc W$-algebra $\mc W(\mf g,f)$. Then, it gives a formula, equation \eqref{eq:Lw}, 
for $L(\partial)$ in terms of the generators $w(u_i),\,i\in I_f$,
of the $\mc W$-algebra $\mc W(\mf g,f)$.

\subsection{Integrable hierarchies of Lax type equation}
Let $\mf g$ be one of the classical Lie algebras $\mf{gl}_N$, $\mf{sl}_N$, $\mf{so}_N$ or $\mf{sp}_N$,
and let $V=\mb F^N$ be its standard representation
(endowed, in the cases of $\mf{so}_N$ and $\mf{sp}_N$,
with a non-degenerate symmetric or skewsymmetric bilinear form, respectively). Then, we can use the 
operator $L(\partial)$ in \eqref{eq:Lw} to get explicit formulas for the $\lambda$-brackets among the generators of $\mc W(\mf g,f)$ and construct integrable hierarchies of Hamiltonian equations, see \cite{DSKVcl}.

\begin{theorem}
\begin{enumerate}[1)]
\item
$L(\partial)$ satisfies the generalized Adler type identity
\begin{equation}\label{eq:adler-general}
\begin{array}{l}
\displaystyle{
\vphantom{\Big(}
\{L(z)_\lambda L(w)\}
=
\alpha
(\id_V\otimes L(w+\lambda+\partial))(z-w-\lambda-\partial)^{-1}
(L^*(\lambda-z)\otimes\id_V)\Omega_V
} \\
\displaystyle{
\vphantom{\Big(}
-\alpha
\Omega_V\,\big(L(z)\otimes(z-w-\lambda-\partial)^{-1}L(w)\big)
} \\
\displaystyle{
\vphantom{\Big(}
-\beta
(\id_V\otimes L(w+\lambda+\partial))
\Omega_V^\dagger(z+w+\partial)^{-1}(L(z)\otimes\id_V)
} \\
\displaystyle{
\vphantom{\Big(}
+\beta
(L^*(\lambda-z)\otimes\id_V)\Omega_V^\dagger(z+w+\partial)^{-1}(\id_V\otimes L(w))
} \\
\displaystyle{
\vphantom{\Big(}
+\gamma\big(\id_V\otimes \big(L(w+\lambda+\partial)-L(w)\big)\big)
(\lambda+\partial)^{-1}
\big(\big(L^*(\lambda-z)-L(z)\big)\otimes\id_V\big)
\,,}
\end{array}
\end{equation}
for the following values of $\alpha,\beta,\gamma\in\mb F$:
$$
\begin{tabular}{l|lll}
\vphantom{\Big(}
$\mf g$ & \,\,$\alpha$\,\, & \,\,$\beta$\,\, & \,\,$\gamma$\,\, \\
\hline 
\vphantom{\Big(}
$\mf{gl}_N$ &  1 & 0 & 0 \\
\vphantom{\Big(}
$\mf{sl}_N$ &  1 & 0 & $\frac1N$ \\
\vphantom{\Big(}
$\mf{so}_N$ or $\mf{sp}_N$ &  $\frac12$ & $\frac12$ & 0 \\
\end{tabular}
$$
In equation \eqref{eq:adler-general} $L^*$ denotes the formal adjoint of pseudodifferential operators,
and 
$\Omega_V$ and $\Omega_V^\dagger$ are defined by equations \eqref{Omega} and
\eqref{Omega-dagger} respectively.
\item
For $B(\partial)$ a $K$-th root of $L(\partial)$ (i.e. $L(\partial)=B(\partial)^K$ for $K\geq1$)
define the elements $h_{n,B}\in\mc W(\mf g,f)$, $n\in\mb Z_\geq0$, by ($\tr=1\otimes\tr$)
$$
h_{n,B}=
\frac{-K}{n}
\Res_z\tr(B^n(z))
\text{ for } n>0
\,,\,\,
h_0=0\,.
$$
Then, all the elements $\tint h_{n,B}$ are Hamiltonian functionals in involution and we have the corresponding integrable hierarchy of Lax type Hamiltonian equations
\begin{equation}\label{eq:hierarchy}
\frac{dL(w)}{dt_{n,B}}
=
\{\tint h_{n,B},L(w)\}
=
[\alpha(B^n)_+-\beta((B^{n})^{*\dagger})_+,L](w)
\,,\,\,n\in\mb Z_{\geq0}\,.
\end{equation}
(In the RHS of \eqref{eq:hierarchy} we are taking the symbol of the commutator of matrix pseudodifferential operators.)
\end{enumerate}
\end{theorem}

\begin{remark}
For $\beta=0$ solutions to the integrable hierarchy \eqref{eq:hierarchy} can be obtained by reductions
of solutions to the multicomponent KP hierarchy, see \cite{last}.
\end{remark}

%
%

\end{document}